\documentclass[review,12pt]{elsarticle}

\usepackage{amssymb}
\usepackage{amsmath}
\usepackage{amsmath,amssymb}
\usepackage{graphicx}
\usepackage{booktabs}
\usepackage{siunitx}
\usepackage{cleveref}   
\usepackage{diagbox}    
\usepackage{xcolor}
\usepackage{subcaption}

\crefname{equation}{Eq.}{Eqs.}
\crefname{figure}{Fig.}{Figs.}
\crefname{table}{Table}{Tables}
\crefname{section}{Section}{Sections}

\journal{Nuclear Instruments and Methods in Physics Research Section B}

\begin{document}
\begin{frontmatter}



\title{Threshold displacement energies and neutron-induced displacement-per-atom response of CsPbBr$_3$ from molecular dynamics and Monte Carlo simulations}


\author[1]{Zhongming Zhang}
\author[1]{Samuel Murphy}
\author[1]{Michael Aspinall}
\affiliation{organization={School of Engineering, Lancaster University},
            addressline={Bailrigg}, 
            city={Lancaster},
            postcode={LA1 4YW}, 
            country={United Kingdom}}

\begin{abstract}
CsPbBr$_3$ is a promising halide perovskite material for ionising-radiation detection, but its displacement-damage response under fast-neutron irradiation remains insufficiently understood. In this work, molecular dynamics simulations were combined with Geant4 Monte Carlo calculations to evaluate threshold displacement energies and neutron-induced displacement-per-atom (DPA) responses in CsPbBr$_3$. An ICSD-based orthorhombic CsPbBr$_3$ structure was used for site-specific threshold displacement energy calculations. Interatomic interactions were described using tabulated Buckingham--ZBL short-range potentials combined with long-range Coulomb interactions, and the potential model was verified by structural relaxation, finite-temperature equilibration and elastic-constant validation against density-functional-theory and experimental reference data. Threshold displacement energies were calculated for Cs, Pb, apical Br and equatorial Br sites over 100 recoil directions at 100, 200 and 300 K, with three independent random seeds for each direction. The results show clear site and directional dependence, with Pb exhibiting the highest average threshold displacement energy and the two nonequivalent Br sites showing distinct displacement responses. The molecular-dynamics-derived threshold displacement energies were then used in Geant4 recoil-damage calculations for fusion-relevant neutron energies of 2.45 MeV and 14.1 MeV. Species-resolved recoil spectra were obtained for Cs, Pb and Br, and DPA values were calculated using a Lindhard-type damage-energy partition and a Norgett--Robinson--Torrens-type displacement model. For a \SI{1}{cm^3} active CsPbBr$_3$ volume, the calculated DPA per incident neutron was $9.056\times10^{-22}$ for 2.45 MeV neutrons and $1.248\times10^{-21}$ for 14.1 MeV neutrons. These results provide atomistic threshold-displacement data and fusion-neutron damage estimates for assessing the radiation tolerance of CsPbBr$_3$ in neutron-detection applications.
\end{abstract}




\end{frontmatter}



\section*{Introduction}

Metal halide perovskites have attracted considerable interest for ionising-radiation detection because of their high average atomic numbers, strong stopping power, wide band gaps, relatively high carrier mobilities and compatibility with low-temperature crystal growth. Among them, CsPbBr$_3$ is an all-inorganic halide perovskite that has been widely investigated for X-ray and gamma-ray detection. CsPbBr$_3$ single crystals and related detector architectures have shown promising performance in X-ray detection and imaging, while recent high-profile studies have further demonstrated the potential of CsPbBr$_3$-based devices for ionising-radiation detection applications \cite{Stoumpos2013,Duan2018,Zhang2019,Pan2020,Zhang2021b,Peng2021CsPbBr3Xray}.

Lead halide perovskites have also begun to be explored for neutron-detection applications. Because most halide perovskites do not contain nuclides with large thermal-neutron capture cross sections, neutron detection is commonly achieved through converter-assisted approaches, in which a neutron-conversion material generates secondary charged particles or photons that are subsequently detected by the perovskite semiconductor. For example, hybrid lead bromide perovskites have been studied with external converter layers for neutron detection, and microstructured CsPbBr$_3$ devices combined with boron converter materials have been reported for thermal-neutron detection \cite{Andricevic2021Neutron,CaraveoFrescas2022CsPbBr3Neutron}. These studies indicate that lead halide perovskites can serve as charge-collection media in neutron-detector concepts.

For detector materials intended for use in neutron-rich or high-flux radiation environments, however, displacement damage is an important concern. Fast neutrons can transfer kinetic energy to lattice atoms through scattering reactions, generating primary knock-on atoms (PKAs). If the transferred recoil energy exceeds the threshold displacement energy, a lattice atom may be displaced from its site, leading to vacancies, interstitials and defect complexes. These defects may modify charge transport, increase carrier trapping and recombination, and degrade detector performance. Therefore, in addition to detector sensitivity and charge-collection properties, the fast-neutron displacement-damage response of CsPbBr$_3$ is an important material question.

A key input for displacement-damage modelling is the threshold displacement energy, $E_d$. This quantity defines the minimum recoil energy required to create a stable displacement defect and enters directly into displacement-per-atom (DPA) calculations. In crystalline materials, $E_d$ is not generally a single universal constant, but depends on the displaced atomic species, crystallographic site, recoil direction and temperature. Molecular dynamics simulations provide a practical route for evaluating such site- and direction-dependent threshold displacement energies and have been widely used in radiation-damage studies of detector, semiconductor and nuclear materials \cite{Buchan2015,He2020a,Gray_2022}.

Recent ab initio molecular dynamics work has begun to quantify threshold displacement energies in lead halide perovskites, including MAPbI$_3$, FAPbI$_3$ and CsPbI$_3$, demonstrating that the displacement thresholds of these soft ionic semiconductors can be considerably lower than values commonly assumed in radiation-damage modelling \cite{MartinezDuque2025}. However, corresponding site-specific and direction-dependent threshold displacement energies for CsPbBr$_3$ remain unavailable. In particular, the two crystallographically nonequivalent Br sites in orthorhombic CsPbBr$_3$ have not been separately evaluated. Moreover, the implications of such atomistic displacement thresholds for neutron-induced recoil spectra and DPA response have not yet been established for CsPbBr$_3$.

In this work, molecular dynamics simulations were first used to calculate site-specific and direction-dependent threshold displacement energies of CsPbBr$_3$ at 100, 200 and 300 K. An ICSD-based orthorhombic structure was employed, and Cs, Pb, apical Br and equatorial Br sites were treated separately. The resulting threshold displacement energies were then incorporated into Geant4 Monte Carlo simulations to evaluate species-resolved recoil spectra and DPA responses. Neutron energies of 2.45 MeV and 14.1 MeV were selected as well-defined fast-neutron benchmark cases corresponding to deuterium--deuterium (D--D) and deuterium--tritium (D--T) fusion neutron energies, respectively. These energies were used not to model a specific reactor deployment scenario, but to provide representative benchmark cases for linking atomistic displacement thresholds to fast-neutron damage metrics in CsPbBr$_3$. The resulting site-resolved thresholds, species-resolved recoil spectra and DPA estimates can support future assessments of the radiation tolerance of CsPbBr3 detectors and provide input parameters for higher-level damage-accumulation or device-performance models.
\section*{Methodology}

\subsection*{Crystal structure and potential}

The atomic models of CsPbBr$_3$ were constructed from an ICSD-based crystallographic information file corresponding to the orthorhombic (\textit{Pbnm}, GdFeO$_3$-type) phase, which is the stable phase over the 100--300~K range considered here. This setting makes the two crystallographically nonequivalent Br sites explicit: the apical (axial) Br1 site (Wyckoff 4c) and the equatorial Br2 site (Wyckoff 8d, in the $ab$ plane). The CIF was converted to LAMMPS input using Atomsk. A $12 \times 12 \times 8$ supercell (23\,040 atoms) was used for the threshold-displacement-energy calculations, and a smaller $5 \times 5 \times 5$ supercell was used for potential verification. Full structural details and the Wyckoff coordinates are given in Supplementary Section~S1 and Supplementary Fig.~1.

Interatomic interactions were described by a tabulated short-range Buckingham--ZBL potential combined with long-range Coulomb interactions evaluated by the PPPM method, with formal ionic charges of $+1$, $+2$ and $-1$ assigned to Cs, Pb and Br, respectively. The equilibrium Buckingham branches follow established lead-halide and alkali-halide parameterisations, while the universal Ziegler--Biersack--Littmark (ZBL) potential governs the close-approach repulsion sampled during recoil events; the two branches are joined by an exponential spline (Supplementary Section~S2). The model was implemented in LAMMPS~\cite{plimpton1995fast} as a hybrid short-range and long-range Coulomb interaction. This is a non-polarisable, formal-charge force field, used here to evaluate \textit{relative} site- and direction-dependent displacement thresholds rather than to reproduce the full electronic or dielectric response of CsPbBr$_3$. The complete parameter set and spline construction are given in Supplementary Section~S2 and Supplementary Table~1.

The model was verified prior to the recoil calculations through 0~K relaxation, 298~K NPT equilibration and an NVE energy-conservation check, all of which left the lattice parameters within approximately 1\% of the reference structure, and through quasi-static drag profiles confirming the smoothness and expected $Z^2$ scaling of the spliced short-range potential. As a quantitative check, the full 0~K elastic-constant tensor was computed and averaged using the Voigt--Reuss--Hill scheme to obtain polycrystalline moduli. The bulk modulus (21.4 GPa) reproduces the DFT-PBE value to within approximately 17\%, whereas the shear and Young's moduli are overestimated by approximately 50\%. The reported threshold displacement energies should therefore be regarded as upper-bound estimates; the implications for the DPA values are discussed in the Limitations section. The full verification procedure and the elastic-constant comparison are detailed in Supplementary Section~S3 (Supplementary Tables~2--3 and Supplementary Fig.~2).

\subsection*{Threshold displacement energy calculations}

Threshold displacement energy calculations were carried out using the Large-scale Atomic/Molecular Massively Parallel Simulator (LAMMPS) \cite{plimpton1995fast}. The threshold displacement energy ($E_d$) is defined here as the minimum recoil energy required to produce a stable defect after relaxation.

The TDE calculations were performed using a $12\times12\times8$ supercell containing 23040 atoms. The simulations were carried out in \texttt{metal} units with periodic boundary conditions in all three directions, and the equilibrated restart file at the target temperature was used as the initial configuration for each recoil trial. The atomic masses were set to 132.90545190, 207.2 and 79.904~amu for Cs, Pb and Br, respectively.

A representative angular sampling grid was used to evaluate the directional dependence of the threshold displacement energy. The polar angle $\theta$ and azimuthal angle $\phi$ were both sampled from $0^\circ$ to $90^\circ$ in steps of $10^\circ$, producing 100 recoil directions for each selected PKA site. This angular range was used as a representative positive crystallographic-direction sampling domain rather than as a strict irreducible wedge of the orthorhombic structure. The reported site-averaged threshold displacement energies are therefore arithmetic averages over the sampled angular grid, not full spherical averages weighted by $\sin\theta$. For the overall summary value reported in Table~\ref{tab:tde}, a stoichiometry- and site-multiplicity-weighted average was calculated using the weighting Cs:Pb:Br1:Br2 = 1:1:1:2, where Br1 and Br2 denote the apical and equatorial Br sites, respectively.

This sampling strategy was chosen to provide a systematic and computationally tractable comparison of the directional displacement response of Cs, Pb, apical Br and equatorial Br sites under the same angular grid. The resulting heat maps should therefore be interpreted as direction-resolved threshold-displacement maps over the sampled angular domain, rather than as complete spherical distributions.

For each recoil trial, the geometric centre of the simulation cell was determined automatically from the box boundaries, and a spherical core region of radius $25.0$~\AA{} was defined as the active recoil region. Atoms outside this core were assigned to a boundary region, which acted as a heat sink during the recoil simulation. The system was evolved using an NVE integrator, while the boundary atoms were coupled to a Langevin thermostat at the target temperature with a damping parameter of 10.0 ps.

For each simulation, the PKA was selected from atoms located close to the centre of the supercell in order to minimise boundary effects. For Cs and Pb, symmetry-equivalent sites near the centre of the simulation box were chosen. For Br, two nonequivalent lattice sites exist in the orthorhombic structure, and these two Br positions were treated separately in the TDE calculations.

At the beginning of each recoil trial, thermal velocities were regenerated at the target temperature using a prescribed random seed and a Gaussian distribution. For each recoil direction, three independent trials with different random seeds were run and the resulting thresholds were averaged. The recoil direction was defined from the selected $\theta$ and $\phi$ values, and the corresponding Cartesian velocity components were calculated from the assigned recoil energy and the mass of the selected PKA. Before applying the recoil velocity to the PKA atom, the total linear momentum of the system was removed.

A variable-timestep scheme was employed during the recoil stage in order to improve numerical stability during close collisions. In the present implementation, the timestep was dynamically adjusted using \texttt{fix dt/reset} between $10^{-5}$~ps and $10^{-3}$~ps. Each recoil simulation consisted of a 25~ps recoil/cascade stage followed by a 10~ps relaxation stage before defect analysis was performed.

The threshold displacement energy for each direction was determined using an automated binary-search workflow implemented in Python. For each task, the temperature, recoil direction, random seed and PKA atom ID were read from an input CSV file. The initial search interval was set to 5--150 eV, and the recoil energy was progressively narrowed by binary search until the interval width became smaller than 1 eV, which was taken as the numerical precision of the TDE evaluation. The resulting threshold energies for all retained valid cases lay within the prescribed search interval, and no retained TDE value was truncated by the 5 eV lower bound or the 150 eV upper bound.

Defect identification was carried out using the Wigner--Seitz analysis method implemented in OVITO \cite{stukowski2009visualization}. A reference frame was written immediately before recoil injection, and the final relaxed configuration after the recoil event was compared against this reference using affine mapping to the reference lattice. The Wigner--Seitz analysis identifies vacant lattice sites and excess occupancies, corresponding to vacancy- and interstitial-type defects in the mapped lattice.

In the automated binary-search workflow, the final vacancy count returned by the Wigner--Seitz analysis was used as the threshold flag. A recoil trial was considered to be above threshold if at least one stable Wigner--Seitz vacancy remained after the recoil and relaxation stages. This criterion was used as a practical indicator of a stable Frenkel-pair-like displacement event, since the creation of a vacancy in the Wigner--Seitz mapping indicates that at least one atom has been displaced from its original lattice site. Trials with no remaining Wigner--Seitz vacancy were considered to be below threshold.

\subsection*{Displacement-per-atom calculation}

The displacement-per-atom (DPA) response of CsPbBr$_3$ under neutron irradiation was evaluated using the Geant4 toolkit \cite{Agostinelli2003,Allison2016}. Monoenergetic neutron energies of 2.45 MeV and 14.1 MeV (the D–D and D–T fusion neutron energies, as motivated in the Introduction) were used. The \texttt{FTFP\_BERT\_HP} physics list was employed, which provides high-precision treatment of neutron interactions in the energy range relevant to the present study.

In the Geant4 model, the CsPbBr$_3$ detector was represented by a cuboid with dimensions of \SI{1.0}{cm} $\times$ \SI{1.0}{cm} $\times$ \SI{1.0}{cm} and a density of \SI{4.42}{g\,cm^{-3}}. A monoenergetic neutron source was defined as a square plane source with dimensions of \SI{1.0}{cm} $\times$ \SI{1.0}{cm}, placed at the front surface of the detector and directed along the $+z$ axis so that the neutrons entered the active volume normally.

\begin{figure}[h]
    \centering
    \includegraphics[width=5.5in]{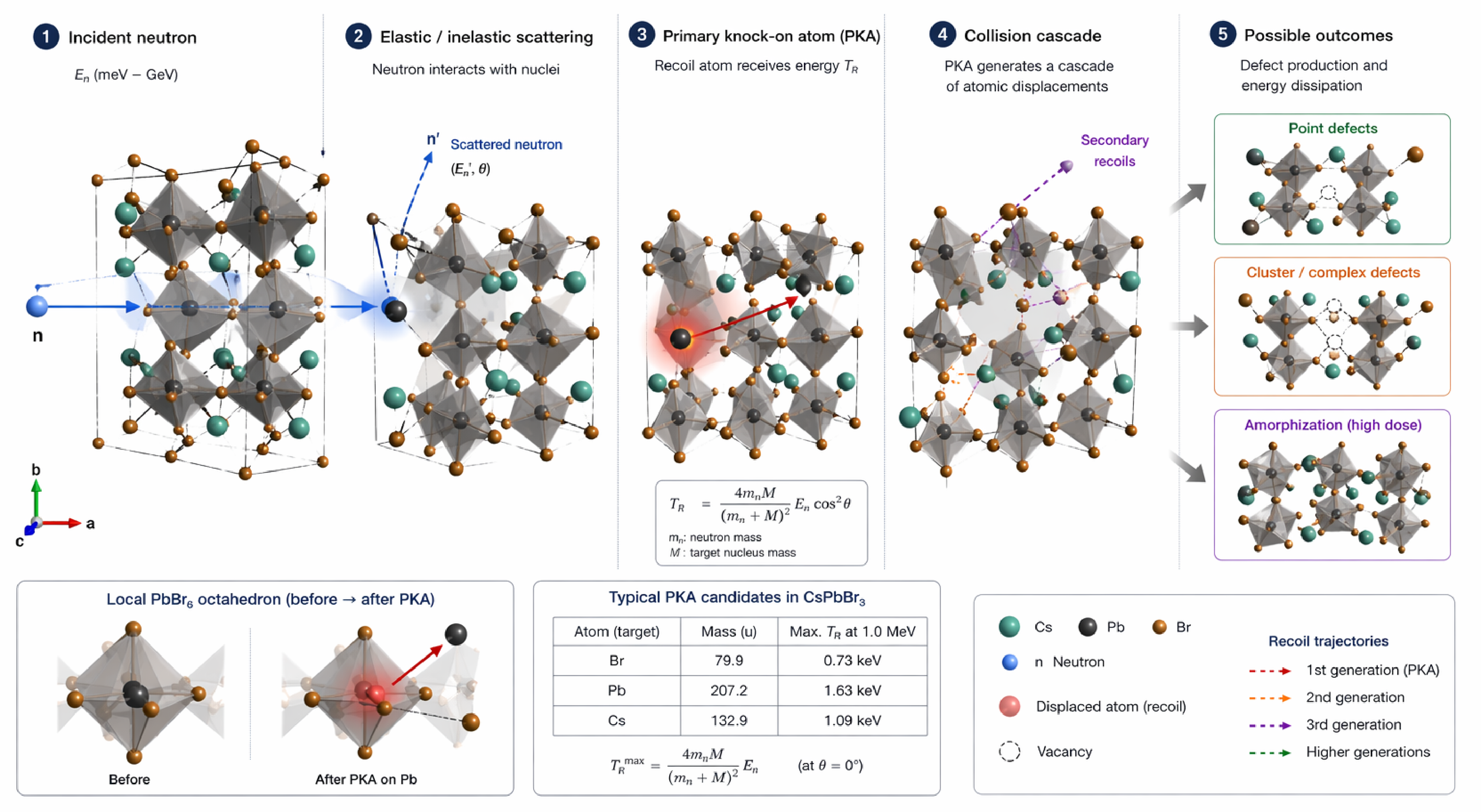}
    \caption{Schematic of neutron-induced displacement processes in CsPbBr$_3$.
Incident neutrons transfer energy to lattice atoms through scattering events, producing PKAs that initiate collision cascades and lead to point defects, defect clusters, or amorphisation.}
    \label{fig:g4}
\end{figure}

The Geant4 simulations were used to record the recoil spectra of Cs, Pb and Br atoms generated by neutron interactions in the active volume. For each recoil atom, the recoil kinetic energy was first converted into the corresponding damage energy using a Lindhard-type partition function, so that only the fraction of the recoil energy available for atomic displacement was retained in the DPA calculation. For each recoil atom, the damage energy $T_{\mathrm{dam}}$ available for atomic displacement was obtained from the recoil kinetic energy $T$ using the Lindhard--Scharff--Schi{\o}tt electronic-stopping partition in the analytic form of Robinson~\cite{Lindhard1963,Robinson1974}, as adopted in the NRT standard~\cite{NRT1975}:
\begin{equation}
T_{\mathrm{dam}} = \frac{T}{1 + k\,g(\varepsilon)}, \qquad g(\varepsilon) = \varepsilon + 0.40244\,\varepsilon^{3/4} + 3.4008\,\varepsilon^{1/6},
\end{equation}
where the reduced energy $\varepsilon$ and the electronic-stopping constant $k$ are given by
\begin{equation}
\varepsilon = \frac{a\,M_2}{Z_1 Z_2 e^2 (M_1 + M_2)}\,T, \qquad k = 0.1337\,Z_1^{1/6}\left(\frac{Z_1}{M_1}\right)^{1/2},
\end{equation}
with $a = 0.8854\,a_0/(Z_1^{2/3} + Z_2^{2/3})^{1/2}$ the screening length and $a_0$ the Bohr radius. Here $Z_1$ and $M_1$ are the atomic number and mass of the recoiling atom, while $Z_2$ and $M_2$ characterise the host medium. Because CsPbBr$_3$ is a multi-component compound in which the recoiling atom slows down in a mixed sublattice environment, the host was represented by its stoichiometric averages $Z_2 = 48.4$ and $M_2 = 115.96$~u, evaluated over the CsPbBr$_3$ formula unit. The partition was applied on a recoil-by-recoil basis within the Geant4 stepping action, prior to evaluating the displacement contribution of Eq.~\ref{eq:nd}.
The corresponding displacement contribution was then evaluated using the Norgett--Robinson--Torrens-type expression\cite{NRT1975}
\begin{equation}
\label{eq:nd}
N_{d}\left(T_{\mathrm{dam}}\right)=
\left\{
\begin{array}{ll}
0, & T_{\mathrm{dam}} < E_{d}, \\
1, & E_{d} \le T_{\mathrm{dam}} < \frac{2E_{d}}{0.8}, \\
\frac{0.8T_{\mathrm{dam}}}{2E_{d}}, & \frac{2E_{d}}{0.8} \le T_{\mathrm{dam}},
\end{array}
\right.
\end{equation}
where $N_d$ is the predicted number of displacements, $T_{\mathrm{dam}}$ is the recoil damage energy available for atomic displacement, and $E_d$ is the threshold displacement energy obtained from the molecular dynamics simulations. In this way, the DPA calculation is based on the nuclear stopping component of the recoil energy rather than on the total recoil kinetic energy.

In the present implementation, the threshold displacement energy was assigned on a recoil-by-recoil basis using the molecular-dynamics-derived directional TDE database. For each recoil atom, the atomic species was identified from its atomic number as Cs, Pb or Br, and the recoil momentum direction was converted into the corresponding polar and azimuthal angles, $(\theta,\phi)$, in the crystal coordinate system. The direction was folded into the first octant using the symmetry-equivalent absolute direction components, and the value of $E_d$ was then obtained by bilinear interpolation on the calculated $(\theta,\phi)$ TDE grid. Thus, the displacement threshold used in Eq.~\ref{eq:nd} was evaluated as
\begin{equation}
E_d = E_d(s,\theta,\phi,T),
\label{eq:ed_directional}
\end{equation}
where $s$ denotes the recoil species and $T$ is the temperature of the TDE dataset. For Br recoils, the two crystallographically nonequivalent Br sites, namely the apical site (Br1) and the equatorial site (Br2), cannot be distinguished in the Geant4 recoil analysis. Therefore, the Br lookup table used in the Geant4 DPA analysis was constructed by combining the molecular-dynamics results from the two Br sublattices at each sampled direction and temperature using their crystallographic site multiplicities. Specifically, an effective Br threshold displacement energy was calculated as
\begin{equation}
    E_{\mathrm{d,Br}}(\theta,\phi,T)=\frac{E_{\mathrm{d,Br1}}(\theta,\phi,T)+2E_{\mathrm{d,Br2}}(\theta,\phi,T)}{3}
\end{equation}
where Br1 and Br2 denote the apical and equatorial Br sites, respectively.

For a given incident neutron history, the total damage contribution was obtained by summing the displacement contributions from all recoil atoms generated in the active volume:
\begin{equation}
D_{\mathrm{event}}=\sum_{i=1}^{N_{\mathrm{recoil}}} N_d(T_{\mathrm{dam},i},E_{d,i}),
\label{eq:damage_event}
\end{equation}
where $N_{\mathrm{recoil}}$ is the number of recoil atoms produced in that neutron history, and $N_d(T_{\mathrm{dam},i},E_{d,i})$ is the displacement contribution associated with the $i$th recoil atom using its species- and direction-dependent threshold displacement energy.

The mean damage per incident neutron was then obtained by averaging over all simulated neutron histories:
\begin{equation}
\overline{D}=\frac{1}{N_{\mathrm{n}}}\sum_{j=1}^{N_{\mathrm{n}}} D_{\mathrm{event},j},
\label{eq:damage_mean}
\end{equation}
where $N_{\mathrm{n}}$ is the total number of incident neutrons simulated.

Finally, the DPA per incident neutron was calculated as
\begin{equation}
\mathrm{DPA}_{\mathrm{n}}=\frac{\overline{D}}{N_{\mathrm{atoms}}},
\label{eq:dpa_per_neutron}
\end{equation}
where $N_{\mathrm{atoms}}$ is the total number of atoms in the active CsPbBr$_3$ volume. In addition to the total DPA, species-resolved damage contributions were also accumulated by separately summing the displacement contributions from Cs, Pb and Br recoil atoms. The corresponding sublattice-normalised DPA values were calculated as
\begin{equation}
\mathrm{DPA}_{s,\mathrm{n}}=
\frac{\overline{D}_{s}}{N_s},
\label{eq:dpa_species}
\end{equation}
where $\overline{D}_{s}$ is the mean displacement contribution per incident neutron from recoils of species $s$, and $N_s$ is the number of atoms of that species in the active volume. For CsPbBr$_3$, the stoichiometric relation $N_{\mathrm{Cs}}=N_{\mathrm{Pb}}=N_{\mathrm{f.u.}}$ and $N_{\mathrm{Br}}=3N_{\mathrm{f.u.}}$ was used.

For a neutron fluence $\Phi$, the corresponding total DPA can then be estimated as
\begin{equation}
\mathrm{DPA}(\Phi)=\mathrm{DPA}_{\mathrm{n}}\,\Phi.
\label{eq:dpa_fluence}
\end{equation}
The same scaling can be applied to the species-resolved DPA values.

In this way, the Geant4 calculations provide species-resolved recoil energy spectra, recoil-direction-dependent TDE sampling, total DPA, and sublattice-normalised DPA responses of CsPbBr$_3$ under the specified neutron irradiation conditions. Statistical convergence with the number of neutron histories and the sensitivity to active-volume thickness were both verified (see Results).

\section*{Results and discussion}

\subsection*{Threshold displacement energies}

\begin{figure}[h]
    \centering
    \includegraphics[width=4.0in]{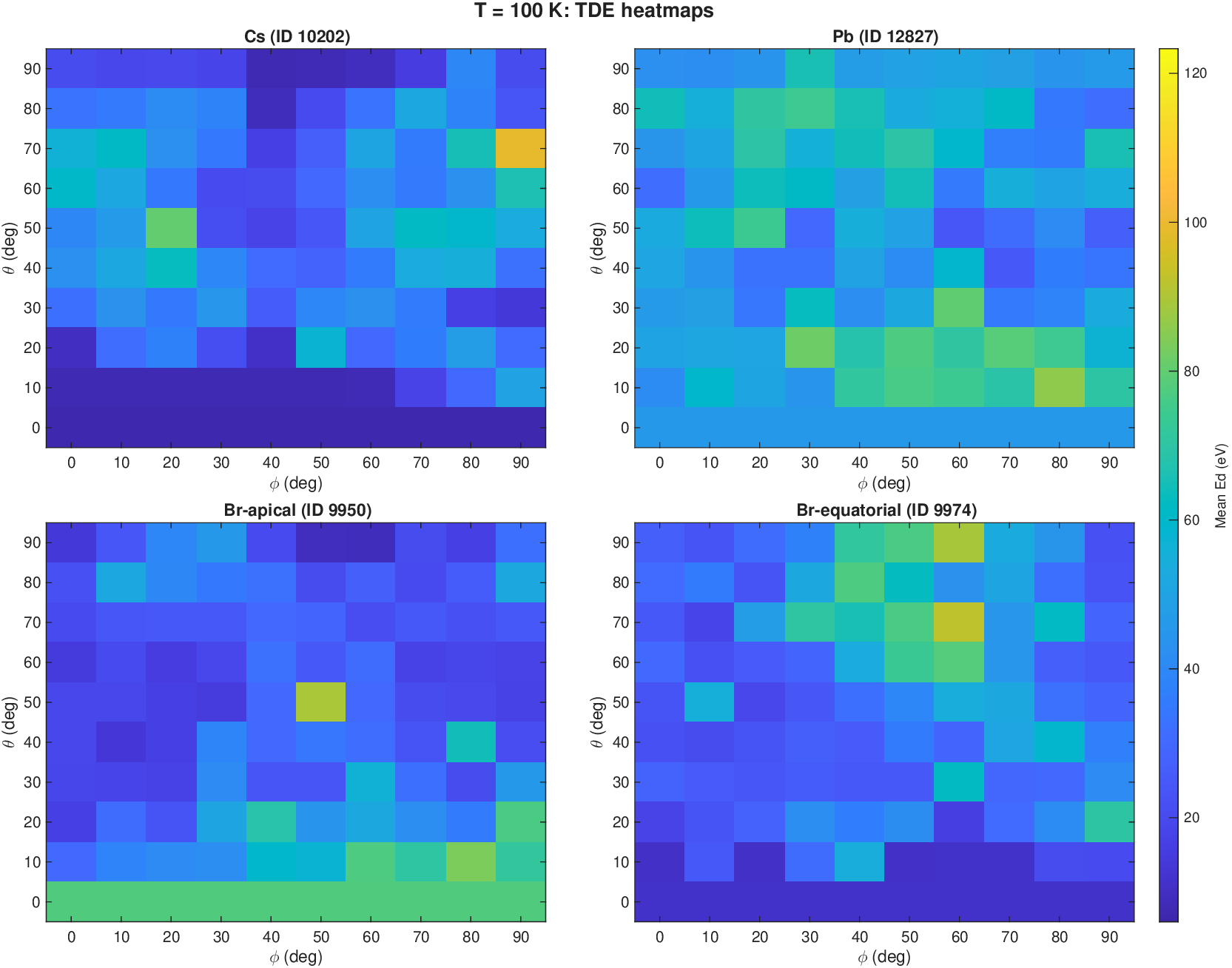}
    \caption{Spatial distribution of threshold displacement energy in CsPbBr$_3$ at 100~K.}
    \label{fig:tde-100k}
\end{figure}

\begin{figure}[h]
    \centering
    \includegraphics[width=4.0in]{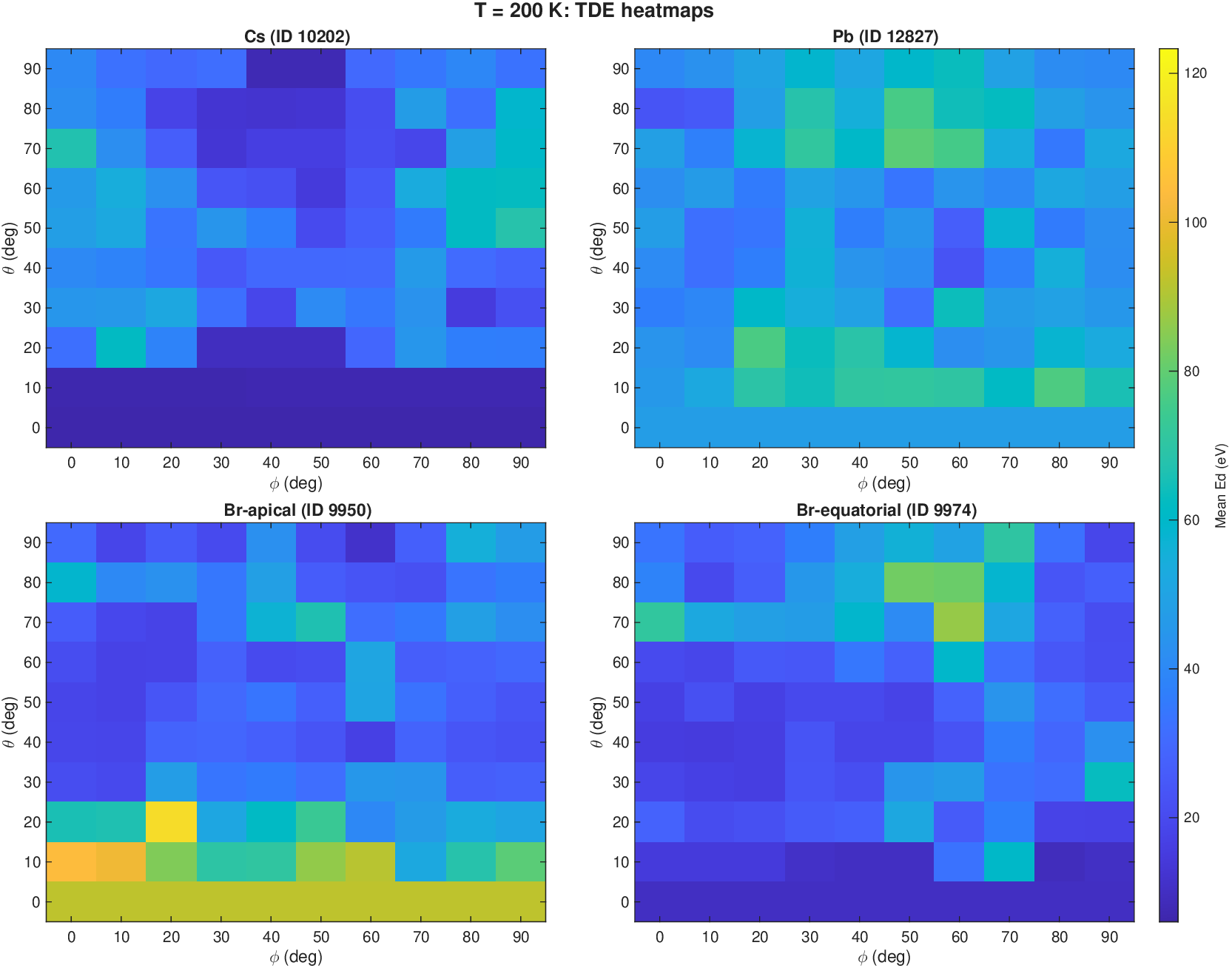}
    \caption{Spatial distribution of threshold displacement energy in CsPbBr$_3$ at 200~K.}
    \label{fig:tde-200k}
\end{figure}

\begin{figure}[h]
    \centering
    \includegraphics[width=4.0in]{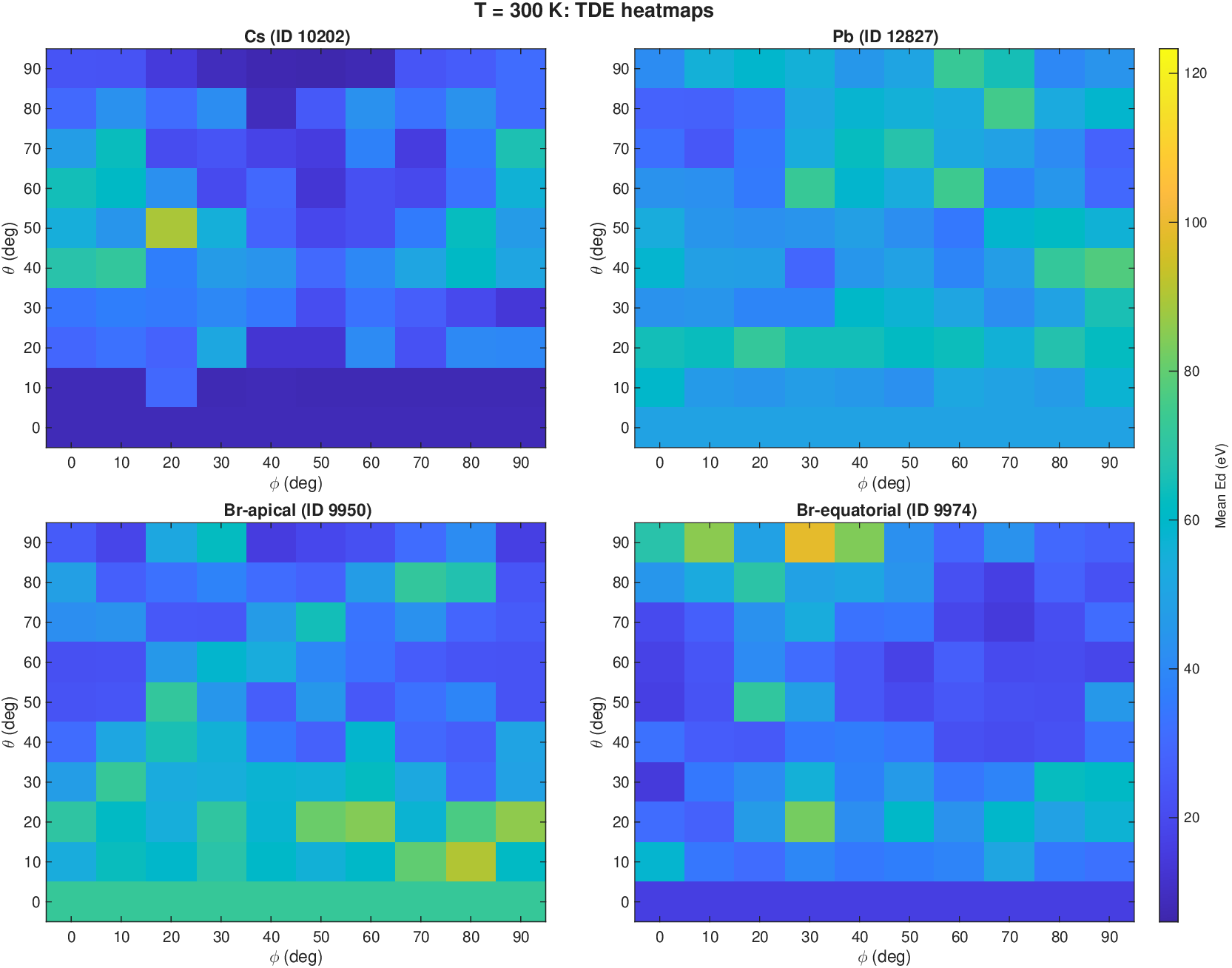}
    \caption{Spatial distribution of threshold displacement energy in CsPbBr$_3$ at 300~K.}
    \label{fig:tde-300k}
\end{figure}

\begin{table}[h]
\centering
\small
\caption{Average threshold displacement energies of CsPbBr$_3$ at different temperatures. Values are given as mean $\pm$ standard deviation from three independent runs with different random seeds. Br1 and Br2 denote the apical and equatorial Br sites, respectively. The weighted average was calculated using the crystallographic site multiplicities Cs:Pb:Br1:Br2 = 1:1:1:2. The unit is eV.}
\label{tab:tde}
\begin{tabular}{@{}lccc@{}}
\toprule
Site & 100 K & 200 K & 300 K \\
\midrule
Cs   & 31.78 $\pm$ 1.45 & 28.99 $\pm$ 0.61 & 29.86 $\pm$ 2.58 \\
Pb   & 52.35 $\pm$ 1.26 & 50.14 $\pm$ 0.69 & 51.09 $\pm$ 1.61 \\
Br1 (apical)     & 37.03 $\pm$ 2.27 & 45.36 $\pm$ 2.27 & 48.68 $\pm$ 1.53 \\
Br2 (equatorial) & 35.50 $\pm$ 3.13 & 30.35 $\pm$ 1.79 & 36.06 $\pm$ 0.60 \\
Weighted average & 38.43 $\pm$ 1.39 & 37.04 $\pm$ 0.87 & 40.35 $\pm$ 0.72 \\
\bottomrule
\end{tabular}
\end{table}

The average threshold displacement energies obtained for the different atomic sites are summarised in \Cref{tab:tde}, while the direction-dependent distributions at 100~K, 200~K and 300~K are shown in \Cref{fig:tde-100k,fig:tde-200k,fig:tde-300k}. The tabulated values provide a compact summary of the site and temperature dependence of the threshold displacement energy, whereas the heat maps highlight the pronounced directional anisotropy of the recoil response.

The results show that the threshold displacement energy in CsPbBr$_3$ is strongly direction-dependent and therefore cannot be represented adequately by a single unique value. For a given atomic site, different crystallographic recoil directions require different minimum recoil energies to produce a stable vacancy--interstitial defect. This behaviour is expected in an ionic crystal, where the local bonding environment, atomic packing and available collision pathways vary with recoil direction.

A clear site dependence is observed in \Cref{tab:tde}. Among the four sites considered, Pb exhibits the highest average threshold displacement energy at all temperatures, with values of 52.35 $\pm$ 1.26~eV, 50.14 $\pm$ 0.69~eV and 51.09 $\pm$ 1.61~eV at 100~K, 200~K and 300~K, respectively. In contrast, Cs shows the lowest average threshold displacement energies, remaining mainly in the range of approximately 28.99--31.78eV. The Br site lies between these two extremes. The difference between Br1 (apical) and Br2 (equatorial) indicates that the two nonequivalent Br lattice sites in the orthorhombic structure have measurably different displacement behaviour.

The site-multiplicity-weighted average threshold displacement energy does not vary monotonically with temperature, taking values of $38.43\pm1.39$, $37.04\pm0.87$ and $40.35\pm0.72$~eV at 100, 200 and 300~K, respectively. This variation (within roughly $\pm3$~eV) is small compared with the site-to-site differences, so the temperature dependence of the average TDE is weaker than its site dependence. One exception is the apical Br1 site, whose average threshold rises monotonically from $37.0$~eV at 100~K to $48.7$~eV at 300~K (Table~1). We do not have a definitive explanation for this trend within the present classical model, and report it as an empirical observation; resolving whether it reflects a temperature-dependent change in the recombination pathways available to displaced apical Br atoms would require dedicated defect-tracking simulations beyond the present scope.

The relatively small standard deviations for most entries in \Cref{tab:tde} indicate that the averaged threshold displacement energies are reasonably reproducible across three independent runs with different random seeds. This suggests that, although the directional distributions are broad, the site-averaged TDE values are statistically stable within the present simulation framework.

Physically, the threshold displacement energy reflects not only the initial ballistic event but also the competition between recombination and defect survival during relaxation, which thermal motion modulates.

These thresholds feed directly into the displacement-per-atom response evaluated in the next section.

\subsection*{Displacement-per-atom response and recoil spectra}

The threshold displacement energies obtained from the molecular dynamics simulations were subsequently combined with Geant4 recoil calculations to evaluate the neutron-induced displacement-per-atom (DPA) response of CsPbBr$_3$. In this section, the statistical convergence of the Geant4 DPA calculation is first assessed, followed by a thickness-sensitivity check of the active scoring volume. The final species-resolved recoil spectra and DPA responses are then compared for 2.45~MeV and 14.1~MeV neutron irradiation.

A neutron-history convergence test was first carried out for the 14.1~MeV case using a \SI{1}{cm^3} active CsPbBr$_3$ volume. As shown in \Cref{tab:dpa-conv}, the calculated DPA per incident neutron stayed within a narrow range, from $1.240\times10^{-21}$ at $10^5$ incident neutrons to $1.248\times10^{-21}$ at $10^6$ incident neutrons. The corresponding mean damage per incident neutron remained close to 29. This indicates that the calculated DPA per incident neutron showed no significant trend as the number of neutron histories increased, and therefore that the 14.1~MeV result had reached statistical convergence. 

\begin{table}[t]
\centering
\small
\setlength{\tabcolsep}{4pt}
\caption{Convergence of the calculated DPA response for 14.1~MeV neutrons in a \SI{1}{cm^3} CsPbBr$_3$ active volume.}
\label{tab:dpa-conv}
\begin{tabular}{@{}cccc@{}}
\toprule
Incident neutrons &
\begin{tabular}[c]{@{}c@{}}Mean damage\\per neutron\end{tabular} &
\begin{tabular}[c]{@{}c@{}}DPA per\\incident neutron\end{tabular} &
\begin{tabular}[c]{@{}c@{}}RMS damage\\per neutron\end{tabular} \\
\midrule
$1\times10^5$ & 28.47 & $1.240\times10^{-21}$ & 230.77 \\
$2\times10^5$ & 29.04 & $1.265\times10^{-21}$ & 235.20 \\
$3\times10^5$ & 29.02 & $1.265\times10^{-21}$ & 231.78 \\
$4\times10^5$ & 28.87 & $1.258\times10^{-21}$ & 229.12 \\
$1\times10^6$ & 28.65 & $1.248\times10^{-21}$ & 232.53 \\
\bottomrule
\end{tabular}
\end{table}

A thickness-sensitivity check was then performed in order to assess the dependence of the volume-averaged DPA metric on the active-volume size. For active volumes of 0.5, 0.6, 0.7, 0.8, 0.9 and \SI{1.0}{cm^3}, using $10^6$ incident neutrons in each case, the calculated DPA per incident neutron remained close to $1.25\times10^{-21}$, as summarised in \Cref{tab:dpa-thickness}. Although the mean damage per incident neutron increased systematically with increasing active volume, from 14.41 at \SI{0.5}{cm^3} to 28.65 at \SI{1.0}{cm^3}, the corresponding DPA per incident neutron varied only weakly. This behaviour is expected because the larger active volume provides a longer interaction path length for the incident neutrons, while the DPA metric is normalised by the total number of atoms in the scoring volume.

\begin{table}[t]
\centering
\small
\setlength{\tabcolsep}{4pt}
\caption{Thickness sensitivity of the calculated DPA response for 14.1~MeV neutrons in CsPbBr$_3$, using $10^6$ incident neutrons in each case.}
\label{tab:dpa-thickness}
\begin{tabular}{@{}cccc@{}}
\toprule
\begin{tabular}[c]{@{}c@{}}Active volume\\(cm$^3$)\end{tabular} &
\begin{tabular}[c]{@{}c@{}}Mean damage\\per neutron\end{tabular} &
\begin{tabular}[c]{@{}c@{}}DPA per\\incident neutron\end{tabular} &
\begin{tabular}[c]{@{}c@{}}RMS damage\\per neutron\end{tabular} \\
\midrule
0.5 & 14.41 & $1.256\times10^{-21}$ & 164.69 \\
0.6 & 17.28 & $1.254\times10^{-21}$ & 180.31 \\
0.7 & 19.98 & $1.244\times10^{-21}$ & 192.47 \\
0.8 & 23.00 & $1.252\times10^{-21}$ & 207.39 \\
0.9 & 25.94 & $1.256\times10^{-21}$ & 222.25 \\
1.0 & 28.65 & $1.248\times10^{-21}$ & 232.53 \\
\bottomrule
\end{tabular}
\end{table}

On this basis, a detector thickness of \SI{1}{cm} was adopted as the reference geometry for the recoil-spectrum and DPA analysis. This choice is physically reasonable because the neutron mean free paths in CsPbBr$_3$ at both 2.45~MeV and 14.1~MeV are substantially larger than \SI{1}{cm}, based on evaluated neutron total cross-section data such as ENDF/B-VIII.0 \cite{Brown2018ENDFB8}. Therefore, the attenuation of the incident neutron flux across the active region is modest, and the neutron field can be regarded as approximately uniform within the scoring volume. The direct Geant4 thickness-sensitivity check further confirms that the predicted DPA per incident neutron is only weakly dependent on the active-volume thickness within the detector-size range considered here. A \SI{1}{cm}-thick active volume was therefore used as a practical reference configuration for extracting recoil spectra and evaluating the volume-averaged DPA response of CsPbBr$_3$.

Using this reference geometry, the final DPA per incident neutron was $1.248\times10^{-21}$ for 14.1~MeV neutrons and $9.056\times10^{-22}$ for 2.45~MeV neutrons, as summarised in \Cref{tab:dpa-energy}. The corresponding mean damage per incident neutron was 28.65 and 20.79, respectively. The 14.1~MeV case therefore produces a stronger displacement response than the 2.45~MeV case.

\begin{table}[t]
\centering
\small
\setlength{\tabcolsep}{4pt}
\caption{Final calculated DPA response of CsPbBr$_3$ under 2.45~MeV and 14.1~MeV neutron irradiation using a \SI{1}{cm^3} active volume and $10^6$ incident neutrons.}
\label{tab:dpa-energy}
\begin{tabular}{@{}cccc@{}}
\toprule
\begin{tabular}[c]{@{}c@{}}Neutron\\energy\end{tabular} &
\begin{tabular}[c]{@{}c@{}}Mean damage\\per neutron\end{tabular} &
\begin{tabular}[c]{@{}c@{}}DPA per\\incident neutron\end{tabular} &
\begin{tabular}[c]{@{}c@{}}RMS damage\\per neutron\end{tabular} \\
\midrule
2.45~MeV & 20.79 & $9.056\times10^{-22}$ & 101.62 \\
14.1~MeV & 28.65 & $1.248\times10^{-21}$ & 232.53 \\
\bottomrule
\end{tabular}
\end{table}

\begin{figure}[h]
    \centering
    \begin{subfigure}[t]{0.48\textwidth}
        \centering
        \includegraphics[width=\textwidth]{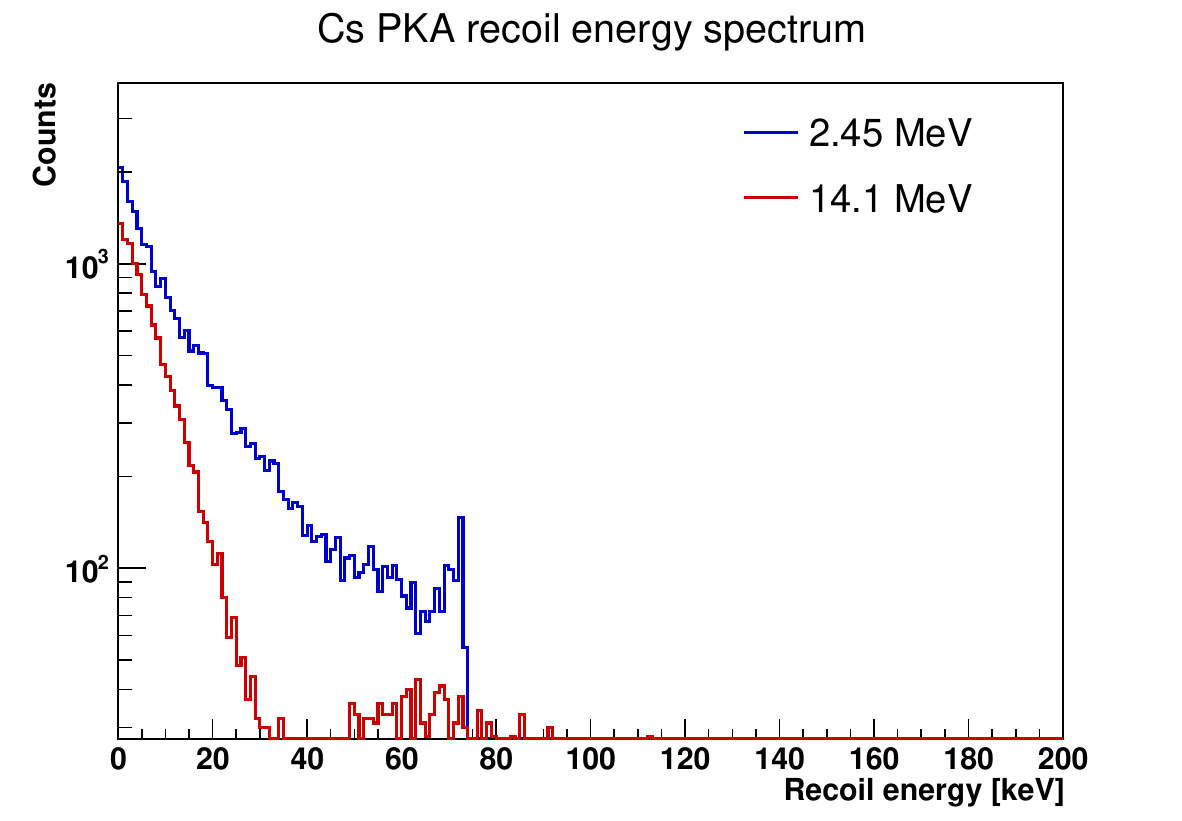}
        \caption{Cs}
        \label{fig:cs-recoil}
    \end{subfigure}
    \hfill
    \begin{subfigure}[t]{0.48\textwidth}
        \centering
        \includegraphics[width=\textwidth]{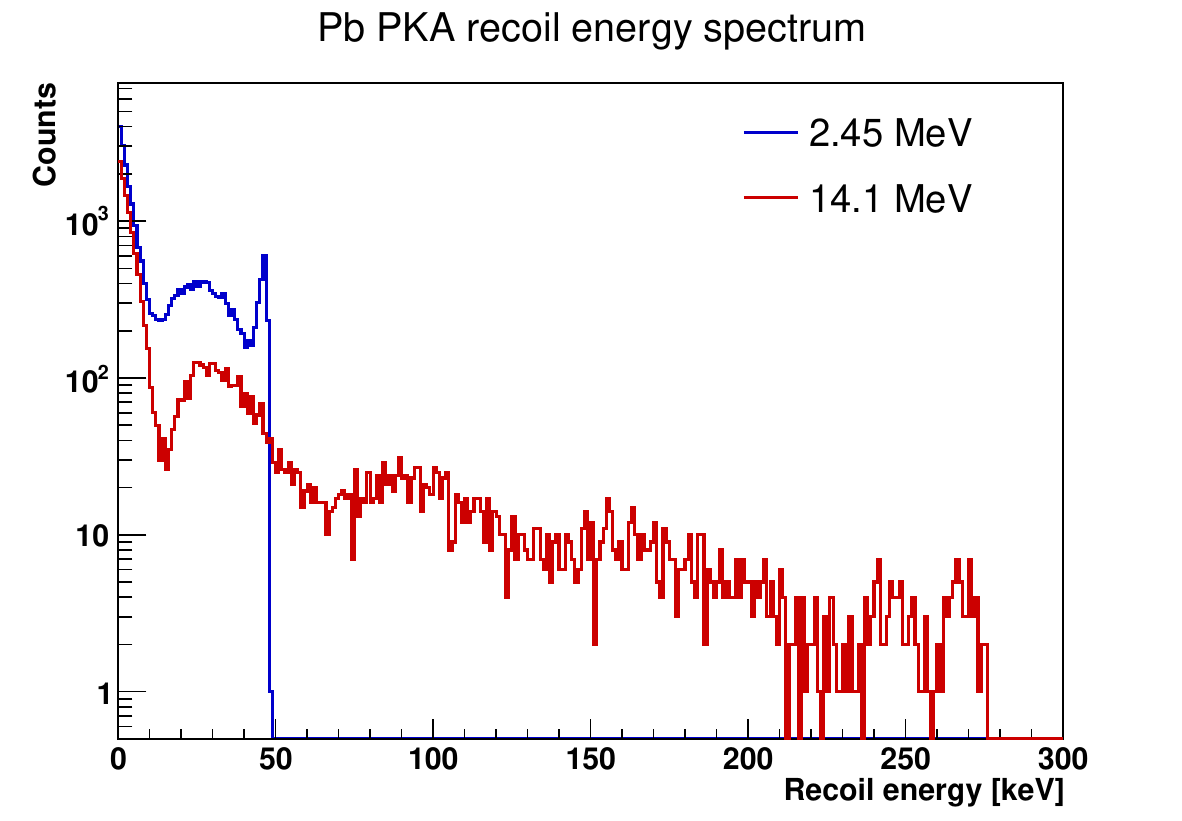}
        \caption{Pb}
        \label{fig:pb-recoil}
    \end{subfigure}

    \vspace{0.5em}

    \begin{subfigure}[t]{0.48\textwidth}
        \centering
        \includegraphics[width=\textwidth]{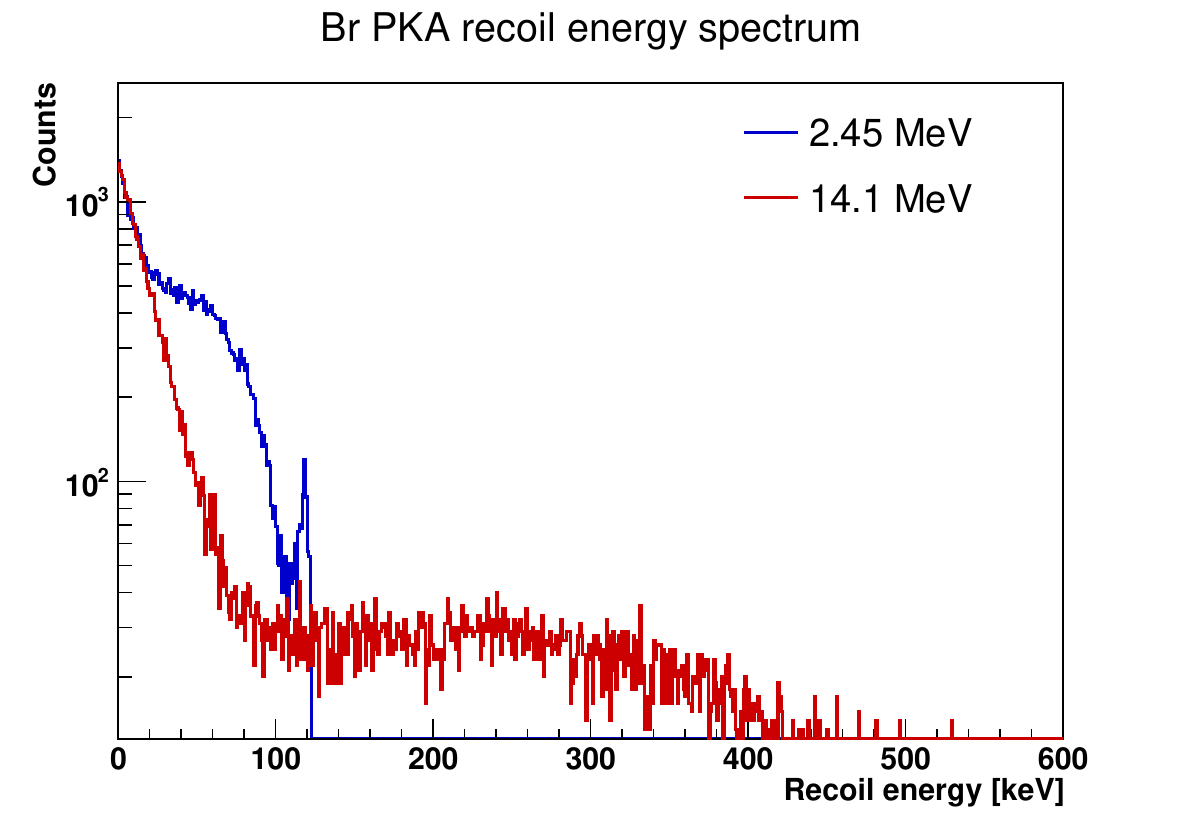}
        \caption{Br}
        \label{fig:br-recoil}
    \end{subfigure}
    \caption{Species-resolved recoil energy spectra of Cs, Pb and Br in CsPbBr$_3$ under neutron irradiation, obtained from the Geant4 simulations. In each panel the blue and red histograms correspond to 2.45~MeV and 14.1~MeV incident neutrons, respectively. The 2.45~MeV spectra show a sharp upper edge set by the maximum elastic recoil energy of each species, while the 14.1~MeV spectra extend to higher recoil energies with more pronounced high-energy tails.}
    \label{fig:recoil-spectra}
\end{figure}

The species-resolved recoil spectra of Cs, Pb and Br are shown in \Cref{fig:recoil-spectra}. For both neutron energies, Br recoils are the most numerous, which is consistent with the stoichiometry of CsPbBr$_3$, where Br occupies three of the five atomic sites per formula unit. At 2.45~MeV, each species shows a sharp upper edge in its recoil spectrum. These edges agree closely with the maximum elastic recoil energy, $E_{\mathrm{R}}^{\max}=4A/(1+A)^2\,E_{\mathrm{n}}$, which gives 119.7~keV for Br, 72.6~keV for Cs and 46.8~keV for Pb at 2.45~MeV, matching the observed cut-offs at approximately 120, 73 and 47~keV. This agreement confirms that the recoil kinematics are correctly reproduced in the Geant4 model. The small peak just below each edge corresponds to near-backscattering elastic collisions, which transfer close to the maximum recoil energy.

At 14.1~MeV, the recoil spectra are broader and extend to higher energies, with a flat high-energy tail that is most pronounced for Pb. This tail arises from inelastic and $(n,xn)$ reaction channels, which become significant for the heavier nuclei at the higher neutron energy. The 2.45~MeV spectra, by contrast, are dominated by elastic scattering and are therefore concentrated below the elastic edge.

The species-resolved DPA values are summarised in \Cref{tab:dpa-species}. For both neutron energies, the Br sublattice accounts for the largest fraction of the total displacements, contributing about 58\% of the summed $N_d$ at 2.45~MeV and about 67\% at 14.1~MeV. When the displacement contributions are instead normalised by the number of atoms of each species, the per-atom DPA is highest for Cs, followed closely by Br, and lowest for Pb. This ordering follows directly from the calculated threshold displacement energies, since Cs has the lowest average $E_d$ and Pb the highest. The Pb sublattice shows almost no change in per-atom DPA between the two neutron energies, because the larger number of low-energy Pb recoils at 2.45~MeV and the smaller number of high-energy Pb recoils at 14.1~MeV give a similar integrated displacement contribution. In contrast, the Br per-atom DPA increases by about 58\% from 2.45 to 14.1~MeV, reflecting the lighter Br mass and the resulting shift of Br recoils to higher energies, where the displacement contribution grows linearly with damage energy.

\begin{table}[t]
\centering
\small
\setlength{\tabcolsep}{5pt}
\caption{Species-resolved DPA per incident neutron in CsPbBr$_3$, normalised by the number of atoms of each species in the active volume (Eq.~\ref{eq:dpa_species}).}
\label{tab:dpa-species}
\begin{tabular}{@{}cccc@{}}
\toprule
Neutron energy & Cs & Pb & Br \\
\midrule
2.45~MeV & $1.373\times10^{-21}$ & $5.158\times10^{-22}$ & $8.797\times10^{-22}$ \\
14.1~MeV & $1.574\times10^{-21}$ & $5.023\times10^{-22}$ & $1.388\times10^{-21}$ \\
\bottomrule
\end{tabular}
\end{table}

The prominence of the Br sublattice in the displacement response is consistent with the known defect chemistry of CsPbBr$_3$. Bromine vacancies are reported to be among the most easily formed native point defects in CsPbBr$_3$, owing to their low formation energy~\cite{Kang2017}, and in bromide-based lead-halide perovskites the deep traps associated with bromide loss can limit charge collection and degrade radiation-detector performance~\cite{Zhao2024}. At the same time, CsPbBr$_3$ is comparatively defect-tolerant---most of its intrinsic defects introduce only shallow levels~\cite{Kang2017}---and single-crystal CsPbBr$_3$ $\gamma$-ray detectors have been shown to retain their spectroscopic performance up to high $^{60}$Co doses~\cite{Zhang2025}, indicating strong tolerance to ionising (gamma) radiation. The present calculation addresses the complementary, ballistic displacement-damage channel: based on recoil kinematics rather than defect formation energies, it provides an independent indication that the Br sublattice is the main site of neutron-induced atomic displacements in CsPbBr$_3$, converging with the formation-energy picture in identifying bromine as the sublattice most active in defect generation under fast-neutron irradiation. It should be noted, however, that in the present model the displacements generated in a recoil cascade are assigned entirely to the sublattice of the primary recoil atom, using that species' threshold displacement energy. The species-resolved DPA values should therefore be interpreted as the displacement contribution initiated by each sublattice, rather than as a complete count of cross-sublattice displacements within the cascade.

Overall, the Geant4 results show that the DPA response of CsPbBr$_3$ is controlled by both the molecular-dynamics-derived threshold displacement energies and the neutron-energy-dependent recoil spectra. The higher DPA response under 14.1~MeV neutron irradiation arises from the broader recoil-energy distributions and the larger recoil damage energies generated in the material, while the species-resolved analysis identifies the Br sublattice as the dominant contributor to neutron-induced displacement damage. These results provide the link between the atomistic threshold displacement calculations and the predicted neutron-damage response of CsPbBr$_3$ under fusion-relevant neutron irradiation conditions.

\subsection*{Estimated neutron fluence required for 1 dpa}

The calculated DPA per incident neutron can be further converted into an estimated neutron fluence required to reach a target displacement dose. From \Cref{eq:dpa_fluence}, the total displacement dose at a neutron fluence $\Phi$ is given by
\begin{equation}
    \mathrm{DPA}(\Phi)=\mathrm{DPA}_{\mathrm{n}}\,\Phi,
    \label{eq:dpa_fluence_result}
\end{equation}
where $\mathrm{DPA}_{\mathrm{n}}$ is the DPA per incident neutron, expressed equivalently as the DPA per unit fluence for the present \SI{1}{cm^2} normally incident beam geometry. Therefore, the neutron fluence required to reach 1 dpa is
\begin{equation}
    \Phi_{1\mathrm{dpa}}=\frac{1}{\mathrm{DPA}_{\mathrm{n}}}.
    \label{eq:fluence_1dpa}
\end{equation}

Using the calculated DPA per incident neutron values, the estimated fluence required to reach 1 dpa is $1.104\times10^{21}$~n\,cm$^{-2}$ for 2.45~MeV neutrons and $8.012\times10^{20}$~n\,cm$^{-2}$ for 14.1~MeV neutrons, as summarised in \Cref{tab:fluence-1dpa}.

\begin{table}[t]
\centering
\small
\setlength{\tabcolsep}{5pt}
\caption{Estimated neutron fluence required to reach 1 dpa in CsPbBr$_3$ under 2.45~MeV and 14.1~MeV neutron irradiation.}
\label{tab:fluence-1dpa}
\begin{tabular}{@{}ccc@{}}
\toprule
Neutron energy & DPA per incident neutron & $\Phi_{1\mathrm{dpa}}$ (n\,cm$^{-2}$) \\
\midrule
2.45~MeV & $9.056\times10^{-22}$ & $1.104\times10^{21}$ \\
14.1~MeV & $1.248\times10^{-21}$ & $8.012\times10^{20}$ \\
\bottomrule
\end{tabular}
\end{table}

If a neutron flux $\varphi$ is specified, the corresponding irradiation time required to reach 1 dpa can be estimated as
\begin{equation}
    t_{1\mathrm{dpa}}=\frac{\Phi_{1\mathrm{dpa}}}{\varphi}.
    \label{eq:time_1dpa}
\end{equation}
For example, for a monoenergetic normally incident neutron flux of $10^{12}$~n\,cm$^{-2}$\,s$^{-1}$, the time required to reach 1 dpa would be approximately 35~years for 2.45~MeV neutrons and 25.4~years for 14.1~MeV neutrons. For a higher flux of $10^{14}$~n\,cm$^{-2}$\,s$^{-1}$, the corresponding times would be approximately 4.2~months and 3.0~months, respectively.

\begin{table}[t]
\centering
\small
\setlength{\tabcolsep}{5pt}
\caption{Illustrative irradiation time required to reach 1 dpa under selected monoenergetic neutron fluxes. These values are model-based displacement-dose estimates and should not be interpreted as detector failure times.}
\label{tab:time-1dpa}
\begin{tabular}{@{}ccc@{}}
\toprule
Neutron flux (n\,cm$^{-2}$\,s$^{-1}$) &
\begin{tabular}[c]{@{}c@{}}2.45~MeV\\time to 1 dpa\end{tabular} &
\begin{tabular}[c]{@{}c@{}}14.1~MeV\\time to 1 dpa\end{tabular} \\
\midrule
$10^{10}$ & 3499~years & 2539~years \\
$10^{12}$ & 35.0~years & 25.4~years \\
$10^{14}$ & 4.2~months & 3.0~months \\
\bottomrule
\end{tabular}
\end{table}

These values should be interpreted as model-based displacement-dose estimates rather than experimentally determined failure thresholds. In particular, the DPA values depend on the threshold displacement energies obtained from molecular dynamics, the Lindhard-type damage-energy partition, and the NRT-type displacement model used in the Geant4 analysis. Moreover, 1 dpa does not necessarily correspond to detector failure; measurable charge-transport degradation may occur at lower displacement doses depending on defect stability, defect migration, carrier trapping and device operating conditions. Nevertheless, the estimated fluence and irradiation-time scales provide useful reference values for comparing the relative displacement-damage response of CsPbBr$_3$ under D--D and D--T fusion-relevant neutron irradiation.

\subsection*{Limitations and future work}
Several limitations of the present modelling framework should be noted. First, the interatomic potential is a non-polarisable formal-charge model. As shown by the elastic-constant validation, it reproduces the bulk modulus of CsPbBr$_3$ to within approximately 17\% of DFT but overestimates the shear and Young's moduli by approximately 50\%, reflecting the absence of electronic polarisability that softens octahedral tilting in halide perovskites. Energy barriers opposing atomic displacement are therefore expected to be overestimated, and the threshold displacement energies reported here should be regarded as upper-bound estimates. This interpretation is consistent with recent ab initio molecular dynamics results for related lead-halide perovskites~\cite{MartinezDuque2025}, which report displacement thresholds considerably lower than values commonly assumed in radiation-damage modelling: the lower bounds of the directional distributions obtained here ($\approx$7--16~eV, Supplementary Table~4) are compatible with that picture, whereas the directional averages are substantially higher, partly because the grid average is dominated by hard directions and partly because of the stiffness of the force field. A direct ab initio evaluation of selected soft directions in CsPbBr$_3$ would provide a valuable benchmark.

Second, the angular sampling covers a single representative atom of each Wyckoff orbit over the first octant. Under the mmm Laue symmetry of the Pbnm structure, the octant is a strict irreducible domain for the orbit-averaged directional threshold displacement energy, which is the quantity relevant to the Geant4 folding procedure; however, the site symmetries of the individual atoms sampled here are lower than mmm, so the single-atom octant maps approximate the orbit averages. The grid averages reported in \Cref{tab:tde} are arithmetic averages over the sampled grid rather than $\sin\theta$-weighted spherical averages. These approximations affect the directional detail of the lookup tables but not the qualitative site ordering, and their influence on the volume-integrated DPA is expected to be limited because the NRT displacement count depends only linearly on the inverse threshold in the high-energy branch.

Third, the displacement model itself carries known limitations. The NRT formalism is known to overestimate the number of stable defects surviving cascade cooling, for which the athermal recombination corrected (arc-dpa) formalism provides an established correction~\cite{Nordlund2018DamageModels}; the present DPA values should therefore be interpreted as standardised exposure metrics rather than absolute surviving-defect counts. In addition, only ballistic displacement damage is considered: ionisation-induced processes such as radiolysis and charge-state-driven defect dynamics, which may be significant in soft ionic semiconductors, are outside the scope of the present framework. The Geant4 mapping of recoil directions into the crystal frame also assumes a single crystal with its crystallographic axes aligned with the beam geometry; polycrystalline or misaligned samples would require an orientational average. Finally, displacements generated within a cascade are assigned to the sublattice of the primary recoil atom, as noted above. A further approximation concerns the effective Br threshold used in the Geant4 lookup (Eq.~(5)), which is a site-multiplicity-weighted linear average of the Br1 and Br2 thresholds formed \emph{before} the NRT displacement function is applied. Because the displacement count in the high-energy branch scales with the inverse threshold, averaging $E_d$ rather than the displacement count underestimates the average of $1/E_d$ (a consequence of Jensen's inequality) and therefore slightly underestimates the Br displacement contribution. The effect is expected to be small given the limited spread between the two Br thresholds, and could be removed in future work by sampling the Br1 and Br2 thresholds with their site probabilities ($1/3$ and $2/3$) on a recoil-by-recoil basis.

Future work will address these points through ab initio validation of selected threshold directions, full collision-cascade simulations to quantify defect survival and clustering beyond the NRT estimate, and the coupling of the present DPA metrics to defect-accumulation and device-performance models. Experimental fast-neutron irradiation of CsPbBr$_3$ detectors, combined with charge-transport characterisation, would provide the most direct test of the damage estimates reported here.
\section*{Conclusions}

In this work, the neutron-induced displacement-damage response of CsPbBr$_3$ was investigated using a combined molecular dynamics and Geant4 simulation approach. An ICSD-based orthorhombic CsPbBr$_3$ structure was used to construct the atomistic models, and the adopted Buckingham--ZBL plus Coulomb interaction model was verified through structural relaxation, finite-temperature equilibration, NVE stability checks and a full elastic-constant validation against DFT and experimental reference data. Because the force field overestimates the shear stiffness, the reported threshold displacement energies are best interpreted as upper-bound estimates. The results confirmed that the potential model provides a stable and elastically reasonable description of the CsPbBr$_3$ lattice for subsequent recoil simulations.

Threshold displacement energies were calculated for Cs, Pb, apical Br and equatorial Br sites at 100, 200 and 300 K. For each selected site and temperature, recoil directions were sampled over a representative positive-direction angular grid using a $10^\circ$ angular step, and three independent random seeds were used to reduce the influence of thermal fluctuations. The results show that the threshold displacement energy of CsPbBr$_3$ is strongly site- and direction-dependent. Pb exhibits the highest average threshold displacement energy while Cs and equatorial Br generally show lower average threshold energies. The difference between the apical and equatorial Br sites further demonstrates that the local crystallographic environment has a measurable influence on the displacement response.

The molecular-dynamics-derived threshold displacement energies were then used in Geant4 recoil-damage calculations for fusion-relevant neutron energies of 2.45~MeV and 14.1~MeV. The Geant4 calculations provided species-resolved recoil spectra for Cs, Pb and Br, which were combined with a Lindhard-type damage-energy partition and a Norgett--Robinson--Torrens-type displacement model to estimate DPA values. Convergence tests showed that the calculated DPA per incident neutron became stable with increasing neutron histories, and thickness-sensitivity checks confirmed that the selected \SI{1}{cm} active volume provides a reasonable reference geometry for the volume-averaged DPA calculation.

For a \SI{1}{cm^3} active CsPbBr$_3$ volume, the calculated DPA per incident neutron was $9.056\times10^{-22}$ for 2.45~MeV neutrons and $1.248\times10^{-21}$ for 14.1~MeV neutrons. The corresponding estimated fluence required to reach 1 dpa is $1.104\times10^{21}$~n\,cm$^{-2}$ for 2.45~MeV neutrons and $8.012\times10^{20}$~n\,cm$^{-2}$ for 14.1~MeV neutrons. The stronger displacement response under 14.1~MeV neutron irradiation is attributed to the broader recoil-energy spectra and higher recoil damage energies produced by the higher-energy neutrons.

Overall, this study provides site-specific threshold displacement energies and fusion-neutron DPA estimates for CsPbBr$_3$. These data are useful for assessing the radiation tolerance of CsPbBr$_3$ as a neutron-detection material and can serve as input parameters for future multiscale simulations of long-term defect accumulation and performance degradation under fast-neutron irradiation.


\section*{Funding}
This work was supported by the School of Engineering, Lancaster University. 

\section*{Conflicts of interest}
The authors declare no conflict of interest.

\appendix





\bibliographystyle{elsarticle-num} 
\bibliography{cas-refs}



\end{document}